\documentclass[12pt]{article}
\usepackage{amssymb}
\usepackage{a4wide}
\usepackage{slashed}
\begin{document}
\pagestyle{empty}
\begin{flushright}
\small
KUL-TF-01/21\\
SISSA 83/01/EP\\
{\bf hep-th/0111130}\\
%June  $16$th, $1998$
\today
\normalsize
\end{flushright}
\begin{center}
{\large\bf Supersymmetry of Massive D=9 Supergravity}\\[.5cm]
{\bf J.~Gheerardyn},$^{\dagger}$\footnote{E-mail: {\tt Jos.Gheerardyn@fys.kuleuven.ac.be}}\footnote{Aspirant
FWO} {\bf P.~Meessen}$^{\dagger\star}$\footnote{E-mail: {\tt meessen@sissa.it}}\\[.2cm]
$^{\dagger}${\em Instituut voor Theoretische Fysica,
Katholieke Universiteit Leuven\\
Celestijnenlaan 200D, B-3001 Leuven, Belgium
}\\[.3cm]
$^{\star}${\em International School for Advanced Studies (SISSA/ISAS)\\
Via Beirut 2-4, 34014 Trieste, Italy}\\[.5cm]
{\bf abstract}\\
\begin{quote}
{\small 
By applying generalized dimensional reduction on the type IIB supersymmetry
variations, we derive the supersymmetry variations for the massive 9-dimensional
supergravity. We use these variations and the ones for massive type IIA to 
derive the supersymmetry transformation of the gravitino for the proposed 
massive 11-dimensional supergravity.
}
\end{quote}
\end{center}
\vspace{.5cm}
%%%%%%%%%%%%%%%%%%%%%%%%%%%%%%%%%%%%%%%%%%%%%%%%%%%%%%%%%%%%%%%%%%%%%%%%%%%
\pagestyle{plain}
%%%%%%%%%%%%%%%%%%%%%%%%%%%%%%%%%%%%%%%%%%%%%%%%%%%%%%%%%%%%%%%%%%%%%%%%%%
The prime example of massive supergravities must be massive type IIA,
or Romans' theory \cite{art:romans}.
As of late it became clear that this theory is dual to type IIA with G-flux,
as can be seen by compactifying massive IIA on a 2-torus \cite{art:singh}. 
Taking into 
account the possible flux, one ends up with an eight dimensional theory,
where the RR mass and the 2-form flux form a vector under the duality
group $Sl(2,\mathbb{R})$,
enabling one to completely rotate the RR mass into flux, which is itself
a manifestation of T-duality between the (massive) IIA and IIB supergravities 
\cite{art:BHO,art:5people}.
Another example of this is the compactification of massive
type IIA on a K3 manifold with fluxes turned on \cite{art:louis,art:bert}.
In that case the RR mass and the fluxes combine into an $O(4,20)$ vector, and the
RR-mass can once again be rotated into flux.
\par
A further connection that was made in the last year is the one between
massive and gauged supergravities. It was shown in \cite{art:koedall} that
the massive $d=9$ theory obtained by Generalized Scherk-Schwarz reduction of IIB supergravity \cite{art:LLP,art:MO},
contained in a certain limit the $N=2$ $d=9$ gauged supergravity. The link
between massive and gauged supergravities was further investigated
in \cite{art:AMO} by considering the dimensional reduction of the bosonic sector of
the proposed
11-dimensional massive supergravity \cite{art:BLO,art:MO} to 8 dimensions, and
showing that the theory automatically has an $SU(2)$ gauge symmetry.
This `gauged supergravity' was not
the same as the one found by Sezgin and Salam \cite{art:SezSal}, though, but seemed
to be related to it by an S-duality transformation. 
A priori this should not be a point of worry: Ungauged $N=2$ $d=8$ supergravity contains two triplets of
vectors, related by $Sl(2,\mathbb{R})$ transformations, and either one
can, in principle, be used to gauge the $SU(2)$ R-symmetry. 
\par
Although we can dualize the type IIA mass-parameter into fluxes, after which the theory
can be oxidized to 11-dimensional supergravity, one would prefer an
action from which to derive massive type IIA without too much effort. In case
of the oxidation of massive IIA such an action was proposed in \cite{art:BLO},
and was afterwards generalized to other theories \cite{art:MO}.
The main feature of these theories is that covariance is `broken' by the
explicit occurrence of killing vectors in the action. This seems necessary since by a well-known no-go theorem
by Deser {\em et. al.} \cite{art:deser} 11-dimensional supergravity cannot
be extended by a cosmological constant, when one assumes covariance.
Since a (killing)vector is dimensionful, one is forced to 
introduce mass-parameters. Some of the parameters are then identified with the charges of the 
M9/KK9-branes. The M9/KK9-brane is, from the point of view of solutions, the oxidation 
of the D8-brane \cite{art:BvdS}. In order to discuss the BPSness of the
M9-brane in the setting of massive 11-dimensional supergravity, a supersymmetry
variation for the gravitino was put forward in \cite{art:BvdS}, which after
dimensional reduction leads to the variations in massive IIA. 
A drawback of the proposed supersymmetry rule is that this variation cannot be straightforwardly generalized
to include more than one killing vector and part of this article is devoted
to its generalization.
\par
The plan of this letter is as follows: In section \ref{sec:IIB} we will
apply generalized dimensional reduction to the type IIB supersymmetry
transformations, in order to obtain the supersymmetry rules for
the 9-dimensional massive supergravity theory presented in \cite{art:LLP,art:MO}.
Using these rules it is discussed that, contrary to the claims made in \cite{art:koedall},
9-dimensional Minkowski space is not a supersymmetric vacuum of the 9-dimensional
gauged supergravity. 
Then in section \ref{sec:M(assive)} we will generalize
the form of the supersymmetry variation of the gravitino in the massive 11-dimensional
supergravity. We will then check that upon dimensional reduction it leads to the correct 
supersymmetry variations for massive type IIA theory and to the supersymmetry rules
derived in Section \ref{sec:IIB}. 
%%%%%%%%%%%%%%%%%%%%%%%%%%%%%%%%%%%%%%%%%%%%%%%%%%%%%%%%%%%%%%%%%%%%%%%%%%
\section{Type IIB SUSY rules and Generalized Dimensional Reduction}
\label{sec:IIB}%%%%%%%%%%%%%%%%%%%%%%%%%%%%%%%%%%%%%%%%%%%%%%%%%%%%%%%%%%%
It is possible to formulate type IIB supergravity in an $SU(1,1)$ covariant way \cite{art:schwarz}. 
The bosonic field content is 2 scalars parameterizing an $SU(1,1)/U(1)$
coset, 
an $SU(1,1)$ doublet of 2-forms $\vec{\mathcal{A}}_{(2)}$, a metric $g$ and 
a 4-form $D_{(4)}$, both transforming as a singlet. The 4-form has self-dual 5-form field strength $F_{(5)}$.
The scalars are combined into an $SU(1,1)$ group matrix $\vec{\mathcal{V}}=(\vec{\mathcal{V}}_- \;
\vec{\mathcal{V}}_+)$ (see the appendix for our choice of parameterization in terms of the usual $Sl(2,\mathbb{R})$ 
scalars).
The $SU(1,1)$ invariant objects are then defined 
as\footnote{In the next sections, we will constantly switch between the $SU(1,1)$ and the
$Sl(2,\mathbb{R})$ formulation. We will use caligraphic letters for objects transforming under the action of
$SU(1,1)$ and use normal letters for their $Sl(2,\mathbb{R})$ counterparts, the only exception being $D_{(4)}$
and its field strength.}
%\footnote{WRT the original article by Schwarz, we rescale 
%$D_{(4)} \to  (4\kappa)^{-1}D_{(4)}$, $A_{(2)}\to \kappa^{-1}A_{(2)}$,
%$\chi\to \kappa^{-1}\chi$ and $\Psi\to \kappa^{-1}\Psi$.
%The same kind of rescalings will be done
%on the M-theory side, in order to get rid of this stupid thind.}
\begin{equation}
  \label{eq:SU11_inv_def}
  \begin{array}{lclclcl}
  \hat{\mathcal{P}}_{(1)} &=& -\hat{\vec{\mathcal{V}}}^{T}_{+}\varepsilon\ 
           d \hat{\vec{\mathcal{V}}}_{+}
     &\hspace{.4cm},\hspace{.4cm}&
  \hat{\mathcal{Q}}_{(1)} &=& -i\hat{\vec{\mathcal{V}}}^{T}_{-}\varepsilon\ 
           d \hat{\vec{\mathcal{V}}}_{+} \, , \\
  \hat{\mathcal{G}}_{(3)} &=& -\hat{\vec{\mathcal{V}}}^{T}_{+}\varepsilon
           \hat{\vec{\mathcal{F}}}_{(3)} &,&
  \hat{\vec{\mathcal{F}}}_{(3)} &=& d\ \hat{\vec{\mathcal{A}}}_{(2)} \; , \\
  \hat{F}_{(5)} &=& d\hat{D}_{(4)}\,+\,\textstyle{\frac{i}{4}}
         \hat{\vec{\mathcal{A}}}_{(2)}\varepsilon\hat{\vec{\mathcal{F}}}_{(3)} &,& & &
  \end{array}
\end{equation}
where $\varepsilon_{12}=+1$. Remark that these objects are not invariant under the (local) $U(1)$,
$\mathcal{Q}_{(1)}$ being the gauge field of this transformation.
In terms of the above objects we can write down the bosonic part of the Non-Self-Dual type IIB action \cite{art:BHO},
\begin{equation}
  \label{eq:SU_NSD_IIB}
  \int_{10}\sqrt{\hat{g}}\left[
             \hat{R}(\hat{g})\,+\,
             2 \hat{\mathcal{P}}^{\hat{\mu}}\hat{\mathcal{P}}_{\hat{\mu}}^{*}\,+\,
            \textstyle{\frac{1}{2\cdot 3!}}
              \hat{\mathcal{G}}_{(3)}^{\hat{\mu}\hat{\nu}\hat{\kappa}}
              \hat{\mathcal{G}}_{(3)\hat{\mu}\hat{\nu}\hat{\kappa}}^{*}
         \,+\,\textstyle{\frac{1}{4\cdot 5!}} \hat{F}_{(5)}^{2}                
                   \right]\,+\,
  \textstyle{\frac{i}{8}}\int_{10} 
       \hat{D}_{(4)}\hat{\vec{\mathcal{F}}}_{(3)}^{T}\varepsilon\hat{\vec{\mathcal{F}}}_{(3)} \; ,
\end{equation}
which, as is usual, has to be supplemented by the self-duality
condition of the 5-form field-strength. In (\ref{eq:SU_NSD_IIB}), we define the components of a p-form to be
\begin{equation}
F_{(p)}=\textstyle{\frac{1}{p!}}F_{\mu_1 \dots \mu_p}dx^{\mu_1} \wedge \dots \wedge dx^{\mu_p}
\end{equation}
\subsection{Ansatz for the bosonic fields}
In regular Kaluza-Klein reduction, one writes the dependence of the higher dimensional fields on the compact
manifold, as a Fourierseries.
Then, reduction means forgetting about all non-zero Fouriercomponents. If one
is reducing on tori (as is the case), the zero components do not have any dependence on the compact
coordinates. However, as IIB supergravity has a global $SU(1,1)$ symmetrie, one can loosen this condition,
giving the `zero modes' a well-defined dependence on the compact coordinate. More concretely, if the
dependence is a specific $SU(1,1)$ transformation, the resulting lower-dimensional theory will not have any
dependence on the circle's coordinate. This is called generalised Scherk-Schwartz reduction
\cite{art:LLP,art:5people,art:MO}.    
Thus, we take the Ansatz for the fields to be
\begin{equation}
  \label{eq:KK_Ansatz}
  \begin{array}{ccc}
  \hat{e}_{\hat{\mu}}{}^{\hat{a}}\;=\; 
     \left(
       \begin{array}{cc}
         K^{3/28}\ e_{\mu}{}^{a} & K^{-3/4}\ A_{(1)\mu} \\
         0 & K^{-3/4}
       \end{array}
     \right) &\hspace{.2cm},\hspace{.2cm} &
     \begin{array}{ccl}
       \hat{D}_{(4)} & =& D_{(4)}\,-\, C_{(3)}\ dy \; ,\\
       \hat{\vec{\mathcal{A}}}_{(2)} &=& \Lambda (y)\left[
                                     \vec{\mathcal{A}}_{(2)}
                                    -\vec{\mathcal{A}}_{(1)}\ dy
                                 \right] \, \\
       \hat{\vec{\mathcal{V}}}_{\pm} &=& e^{\pm i\Sigma}\ \Lambda (y)\ \vec{\mathcal{V}}_{\pm} \; ,
     \end{array}
  \end{array}
\end{equation}
where the $y$ dependence is made explicit. The $\Lambda (y)$ is an $SU(1,1)$ matrix which depends on $y$ in such
a way that $M\equiv \Lambda^{-1}\partial_{y}\Lambda$ is a constant
matrix, which we take to be\footnote{ In this form the masses have the same labeling as in \cite{art:MO}.
Therefore Romans' is recovered when $m_{3}=m_{2}$, $m_{1}=0$ and 
Cowdall's gauged sugra when $m_{1}=m_{2}=0$, $m_{3}\neq 0$.}
\begin{equation}
  \label{eq:Def_mass_matrix}
  M \;=\; \textstyle{\frac{1}{2}}\left(
            \begin{array}{cc}
               i\ m_{3} & i\ m_{2} -m_{1} \\
               -i\ m_{2}-m_{1} & -i\ m_{3}
            \end{array}
          \right) \; .
\end{equation}
The matrix $\Lambda$ corresponding to the mass-matrix in Eq. (\ref{eq:Def_mass_matrix}) is determined by
\begin{equation}
 u\;=\; \cosh\left(\alpha y\right) \,+\, i\textstyle{\frac{m_{3}}{2\alpha}}\sinh\left(\alpha y\right) \;\;,\;\;
 v\;=\; -\textstyle{\frac{m_{1}-im_{2}}{2\alpha}}\sinh\left(\alpha y\right) \; ,
\end{equation}
where $4\alpha^{2}=m_{1}^{2}+m_{2}^{2}-m_{3}^{2}$. (See the Appendix for the meaning of $u$ and $v$.)
$\Sigma$ is the parameter for the necessary\footnote{It can be seen that this transformation is necessary whenever $m_{2}\neq m_{3}$.} 
compensating $U(1)$-transformation, its form is determined through Eq. (\ref{eq:Comp_Trans}).
\par
With the above KK Ansatz we can derive
\begin{displaymath}
  \label{eq:Red_bos_objects}
  \begin{array}{lclclcl}
    \hat{\mathcal{P}}_{a} &=& e^{2i\Sigma}\ K^{-3/28}\ \mathcal{P}_{a} &\; :\;&
    \mathcal{P} &=& -\vec{\mathcal{V}}_{+}^{T}\varepsilon\mathcal{D}\vec{\mathcal{V}}_{+} \; ,\\
    && && && \\
    \hat{\mathcal{Q}}_{a} &=& K^{-3/28} \mathcal{Q}_{a}+\hat{e}_{a}{}^{\hat{\mu}}\partial_{\hat{\mu}}\Sigma &:&
    \mathcal{Q}_{(1)} &=& -i\ \vec{\mathcal{V}}_{-}^{T}\varepsilon\mathcal{D}\vec{\mathcal{V}}_{+} \; ,\\ 
    && && && \\
    \hat{\mathcal{P}}_{9} &=& -K^{3/4}\ \vec{\mathcal{V}}_{+}^{T}\
                                \varepsilon M\vec{\mathcal{V}}_{+} &:&
    \hat{\mathcal{Q}}_{9} &=& -i K^{3/4}\ \vec{\mathcal{V}}_{-}^{T}\ 
                                \varepsilon M\vec{\mathcal{V}}_{+} +
                                 K^{3/4}\partial_{y}\Sigma \; , \\
    && && && \\
    \hat{\vec{\mathcal{G}}}_{(3)abc} &=& e^{i\Sigma}K^{-9/28}\ \vec{\mathcal{G}}_{(3)abc} &\; :\;&
    \vec{\mathcal{F}}_{(3)} &=& d\vec{\mathcal{A}}_{(2)}
    \,+\, A_{(1)}\vec{\mathcal{F}}_{(2)} \; ,\\
    && && && \\
    \hat{\vec{\mathcal{G}}}_{(3)ab9} &=& -e^{i\Sigma}K^{15/28}\  \vec{\mathcal{G}}_{(2)ab} &:&
    \vec{\mathcal{F}}_{(2)} &=& d\vec{\mathcal{A}}_{(1)}\,-\, M\vec{\mathcal{A}}_{(2)} \; ,\\
    && && && \\
    \hat{F}_{(5)a_{1}\ldots a_{5}} &=& K^{-15/28}F_{(5)a_{1}\ldots a_{5}} &:&
    \hat{F}_{(5)a_{1}\ldots a_{4}9} &=& -K^{9/28}F_{(4)a_{1}\ldots a_{4}} \; , 
  \end{array}
\end{displaymath}
\begin{eqnarray}  
    F_{(5)} &=& dD_{(4)} + A_{(1)}F_{(4)}
        +\textstyle{\frac{i}{4}} A_{(1)}
           \vec{\mathcal{A}}_{(4)}^{T}\varepsilon\vec{\mathcal{F}}_{(2)}+
        \textstyle{\frac{i}{4}}\vec{\mathcal{A}}_{(2)}^{T}\varepsilon\vec{\mathcal{F}}_{(3)} \;
        ,\nonumber\\&&\nonumber\\
    F_{(4)} &=& dC_{(3)}-
        \textstyle{\frac{i}{4}}\vec{\mathcal{A}}_{(1)}^{T}\varepsilon\vec{\mathcal{F}}_{(3)}-
        \textstyle{\frac{i}{4}}\vec{\mathcal{A}}_{(2)}^{T}\varepsilon\vec{\mathcal{F}}_{(2)}+
        \textstyle{\frac{i}{4}}A_{(1)}\vec{\mathcal{A}}_{(1)}^{T}
           \varepsilon\vec{\mathcal{F}}_{(2)} \; , 
  \end{eqnarray}
where we have defined
$\mathcal{D}\vec{\mathcal{V}}=(d-A_{(1)}M)\vec{\mathcal{V}}$ and
$\mathcal{G}_{(n)}=-\vec{\mathcal{V}}_{+}^{T}\epsilon\vec{\mathcal{F}}_{(n)}$.
Since the reduced action is, after rewriting it in terms of $Sl(2,\mathbb{R})$
objects, the same as the one presented in \cite{art:MO}, we abstain
from presenting it here. 
%%%%%%%%%%%%%%%%%%%%%%%%%%%%%%%%%%%%%%%%%%%%%%%%%%%%%%%%%%%%%%%%%%%%%%%%%%
\par
%%%%%%%%%%%%%%%%%%%%%%%%%%%%%%%%%%%%%%%%%%%%%%%%%%%%%%%%%%%%%%%%%%%%%%%%%%
As is known, the reduced theory is invariant under local $U(1)$
transformations. In the $SU(1,1)$ notation the action of these transformations on the 
bosonic fields are 
\begin{equation}
\begin{array}{lclclcl}
\vec{\mathcal{V}}_{\pm} &\rightarrow& e^{\pm i\Omega}\ e^{\aleph M}\vec{\mathcal{V}}_{\pm} &\hspace{.3cm},\hspace{.3cm}&
A_{(1)} &\rightarrow& A_{(1)}\,+\, d\aleph \; ,\nonumber \\
\vec{\mathcal{A}}_{(1)} &\rightarrow& e^{\aleph M}\vec{\mathcal{A}}_{(1)} &,&
\vec{\mathcal{A}}_{(2)} &\rightarrow& e^{\aleph M}\left[
                                \vec{\mathcal{A}}_{(2)}+
                                d\aleph\wedge\vec{\mathcal{A}}_{(1)}
                            \right] \; , 
\end{array}\label{eq:9d_bos_local_sym}
\end{equation} 
where $\aleph$ is an arbitrary function and the rest of the objects are invariant.
Once again we stress that we need to apply a compensating transformation, with a parameter
$\Omega$ which is determined by the form of the $SU(1,1)$-matrix $e^{\aleph M}$,
in order to stay in the chosen parameterization of the coset.
%%%%%%%%%%%%%%%%%%%%%%%%%%%%%%%%%%%%%%%%%%%%%%%%%%%%%%%%%%%%%%%%%%%%%%%%%%
\subsection{Reduction of the SUSY transformations}
\label{sec:9d_susy}
%\par
%%%%%%%%%%%%%%%%%%%%%%%%%%%%%%%%%%%%%%%%%%%%%%%%%%%%%%%%%%%%%%%%%%%%%%%%%%
The fermionic field content of the theory consists of a chiral complex Rarita-Schwinger
field $\Psi_{\mu}$ and an anti-chiral complex spinor $\chi$. 
The chiralities of the fermions in the theory are
\begin{equation}
  \label{eq:IIB_Chirality}
  \Gamma_{11}\hat{\Psi}_{\hat{\mu}} \;=\; -\hat{\Psi}_{\hat{\mu}} \hspace{.3cm},\hspace{.3cm}
%  \Gamma_{11}\epsilon \;=\; -\epsilon \hspace{.3cm},\hspace{.3cm}
  \Gamma_{11}\hat{\chi} \;=\; \hat{\chi} \; .
\end{equation}
The fermions are $SU(1,1)$ singlets but transform under the local $U(1)$ with weight $1/2$ resp. $3/2$.
Since we have fields that are charged under the local $U(1)$ we 
define the $U(1)$ covariantized derivative $\mathfrak{D}$ by
\begin{equation}
  \Phi^{[U]}\to e^{iU\aleph (x)}\Phi^{[U]}
    \hspace{.3cm}\longrightarrow\hspace{.3cm}
  \mathfrak{D}\Phi^{[U]}\,\equiv\, \nabla\Phi^{[U]}
                         \,-\, iU\mathcal{Q}_{(1)}\Phi^{[U]}
\end{equation}
Note that $\mathcal{Q}_{(1)}$ acts as a connection for the local 
$U(1)$ since it transforms as $\mathcal{Q}_{(1)}\to\mathcal{Q}_{(1)}
+d\aleph$.
\par
The type IIB supersymmetry transformations are \cite{art:schwarz,art:HW} 
\begin{eqnarray}
  \label{eq:susy_schwarz}
%  \delta \hat{e}_{\hat{\mu}}{}^{\hat{a}} &=&
%        -2 Im \left( \bar{\hat{\epsilon}}\Gamma^{\hat{a}}\hat{\Psi}_{\hat{\mu}} \right) 
%    \; , \nonumber \\
%
  \delta\ \hat{\Psi}_{\hat{\mu}} &=& 
    \hat{\mathfrak{D}}_{\hat{\mu}}\hat{\epsilon} \,+\,
    \textstyle{\frac{i}{16\cdot 5!}} 
          \hat{\slashed{F}}_{(5)}\hat{\Gamma}_{\hat{\mu}}\hat{\epsilon} \,+\,
    \textstyle{\frac{1}{96}}\left[
             \hat{\Gamma}_{\hat{\mu}}{}^{\hat{\nu}\hat{\rho}\hat{\sigma}}
                \hat{\mathcal{G}}_{\hat{\nu}\hat{\rho}\hat{\sigma}}
             -9 \hat{\Gamma}^{\hat{\rho}\hat{\sigma}}
                \hat{\mathcal{G}}_{\hat{\mu}\hat{\rho}\hat{\sigma}}
          \right]\epsilon^{*} \; , \\
  \delta\ \hat{\chi} &=& i\hat{\slashed{\mathcal{P}}}\hat{\epsilon}^{*}
      \,-\, \textstyle{\frac{i}{4\cdot 3!}}
              \hat{\slashed{\mathcal{G}}}_{(3)}\hat{\epsilon} \; , \nonumber 
\end{eqnarray}
where
\begin{equation}
  \hat{\mathfrak{D}}_{\hat{\mu}}\hat{\epsilon} \;=\; 
     \partial_{\hat{\mu}}\hat{\epsilon}\,-\,
     \textstyle{\frac{1}{4}}
          \hat{\slashed{\omega}}_{\hat{\mu}}\hat{\epsilon}\,-\,
     \textstyle{\frac{i}{2}}\hat{\mathcal{Q}}_{\hat{\mu}}\hat{\epsilon} \; .
\end{equation}
since $\hat{\epsilon}$ is a $U(1)$ charge $1/2$ complex Weyl spinor \cite{art:schwarz}. The chirality
of $\hat{\epsilon}$ is defined by $\Gamma_{11}\hat{\epsilon} =-\hat{\epsilon}$
and $\hat{\epsilon}^*$ is the complex conjugate of
$\hat{\epsilon}$.
In order to reduce the susy transformations we decompose the gamma matrices as
\begin{equation}
  \label{eq:IIB_spinor_red}
  \Gamma_{11} \,=\,\sigma^{3}\otimes\mathbb{I}\;,\;
  \Gamma^{9}  \,=\, -i\sigma^{1}\otimes\mathbb{I}\;,\;
  \Gamma^{a}  \,=\, \sigma^{2}\otimes \gamma^{a} \; ,
\end{equation}
where the $\gamma$'s are matrices forming a real Majorana representation\footnote{The signature of the tangent space metric is taken to be
mostly minus.}
of the 9-dimensional Clifford algebra with 
$\gamma^{8}=\gamma^{0}\ldots\gamma^{7}$.
The 10-dimensional chiral spinors reduce to ordinary spinors in 9 dimensions,
{\it i.e.} to a complex combination of 2 Majorana spinors. 
\par
The KK-Ansatz for the 10-dimensional spinors is
\begin{equation}
  \label{eq:10d-gravitino}
  \begin{array}{lclclcl}
  \hat{\chi} & = & e^{3i\Sigma /2}K^{-3/56}\chi &\hspace{.3cm},\hspace{.3cm}&
  \hat{\Psi}_{a} & = & e^{i\Sigma /2}K^{-3/56} 
     \left( \Psi_{a} \,-\,
        \textstyle{\frac{1}{7}}\Gamma_{a}\Gamma^{9}\tilde{\chi}
     \right) \; , \\
  && && && \\
  \hat{\epsilon} &=& e^{i\Sigma /2}K^{3/56} \epsilon &,&
  \hat{\Psi}_{9} &=& e^{i\Sigma /2}K^{-3/56} \tilde{\chi} \; .
  \end{array}
\end{equation}
Please note that due
to the local $U(1)$-invariance we are forced to make this Ansatz for the fermions.
It is also the same local invariance that guarantees that the final result will be,
and indeed is, independent of $\Sigma$. Also note that the above Ansatz implies 
some well determined $y$-dependence of the supersymmetry transformations parameters.
\par
Straightforward reduction of the gravitino and the dilatino variations (\ref{eq:susy_schwarz})
then gives rise to the 9-dimensional supersymmetry variations (the fermionic fields are set to zero):
\begin{eqnarray}
  \delta\chi &=&
    \slashed{\mathcal{P}}\epsilon^{*}
     \,-\,
    K^{6/7}\ \vec{\mathcal{V}}_{+}^{T}\varepsilon M\vec{\mathcal{V}}_{+}\ \epsilon^{*}
     \,+\,
    \textstyle{\frac{1}{8}}K^{9/14}\slashed{\mathcal{G}}_{(2)}\epsilon
     \,-\,
    \textstyle{\frac{1}{4\cdot 3!}}K^{-3/14}
         \slashed{\mathcal{G}}_{(3)} \epsilon \; , \label{eq:9d_trafo_dila}\\
%%%%%
 & & \nonumber \\
%%%%%
  \delta\tilde{\chi} &=&
     -\textstyle{\frac{3}{8}}\slashed{\partial}\log\left( K\right)\epsilon
      \,+\,
      \textstyle{\frac{1}{8}}K^{-6/7}\ \slashed{F}_{(2)}\epsilon
      \,-\,
      \textstyle{\frac{1}{2}}K^{6/7}\
            \vec{\mathcal{V}}_{-}^{T}\varepsilon M\vec{\mathcal{V}}_{+}\ \epsilon
      \,-\,
      \textstyle{\frac{i}{8\cdot 4!}}K^{3/7}\
         \slashed{F}_{(4)}\ \epsilon \nonumber \\
  & & \nonumber \\
  & & +\textstyle{\frac{1}{96}} K^{-3/14}
          \slashed{\mathcal{G}}_{(3)}\ \epsilon^{*} 
      \,+\,
      \textstyle{\frac{3}{32}}K^{9/14}
          \slashed{\mathcal{G}}_{(2)}\ \epsilon^{*} \; , \label{eq:9d_trafo_kkdila}\\
%%%%%%
 & & \nonumber \\
%%%%%%
\delta \Psi_{\mu} &=&
  \mathfrak{D}_{\mu}\epsilon
   \,-\,
  \textstyle{\frac{1}{14}}K^{6/7}\ \vec{\mathcal{V}}_{-}^{T}\varepsilon M\vec{\mathcal{V}}_{+}\
     \gamma_{\mu}\epsilon
   \,+\,
  \textstyle{\frac{1}{7\cdot 8}}K^{-6/7}F_{(2)ab}
     \left[
        \gamma_{\mu}{}^{ab}- 12e_{\mu}{}^{a}\gamma^{b}
     \right] \epsilon \nonumber \\
  & & \nonumber \\
  & & -
  \textstyle{\frac{1}{7\cdot 8}}K^{9/14}\mathcal{G}_{(2)ab}
     \left[
        \gamma_{\mu}{}^{ab}- 12 e_{\mu}{}^{a}\gamma^{b}
     \right] \epsilon^{*}
   \,+\,
  \textstyle{\frac{1}{4\cdot 7\cdot 3!}}K^{-3/14}\mathcal{G}_{(3)abc}
     \left[  
        2\gamma_{\mu}{}^{abc}- 15e_{\mu}{}^{a}\gamma^{bc}
     \right] \epsilon^{*} \nonumber \\
  & & \nonumber \\
  & & +
  \textstyle{\frac{i}{4\cdot 7\cdot 4!}} F_{(4)abcd}
    \left[
       3\gamma_{\mu}{}^{abcd} \,-\, 16 e_{\mu}{}^{a}\gamma^{bcd}
    \right]\epsilon \; , \label{eq:9d_trafo_gravi}
\end{eqnarray}
Note that we still have the local $U(1)$ symmetry and that the local symmetry
relations of Eq. (\ref{eq:9d_bos_local_sym}) are extended to the supersymmetry variation
once we take into account the necessary compensating $U(1)$ transformations.
%%%%%%%%%%%%%%%%%%%%%%%%%%%%%%%%%%%%%%%%%%%%%%%%%%%%%%%%%%%%%%%%%%%%%%%%%%%
%\subsection{Upstairs-downstairs supersymmetry}
\par
%%%%%%%%%%%%%%%%%%%%%%%%%%%%%%%%%%%%%%%%%%%%%%%%%%%%%%%%%%%%%%%%%%%%%%%%%%%
In order to deduce the transformation rules for $N=2$ gauged supergravity, one has to set $m_1=m_2=0$ in the
mass-matrix $M$ \cite{art:koedall}.
In this reference, it was claimed that 10-dimensional
Minkowski space survived the dimensional reduction, leading to
the most uncommon of occurrences that 9 dimensional Minkowski
space was a supersymmetric vacuum of 9-dimensional gauged
supergravity. 
One can check however, by setting $K=1$, $g_{\mu\nu}=\eta_{\mu\nu}$ and all other fields to zero while taking e.g.
$i\gamma^8 \epsilon=\epsilon$, that the supersymmetry
variations Eqs. (\ref{eq:9d_trafo_dila},\ref{eq:9d_trafo_kkdila},\ref{eq:9d_trafo_gravi})
do not support this claim.\footnote{On a more technical side, we remark that the
parameter $c$ in the Bogomol'nyi equations stated in \cite{art:koedall}
has to be set to zero. This can be found using the above supersymmetry rules for a domain-wall solution.
The conclusions about 9-dimensional Minkowski
space, though, were drawn using $c=-1$.} 
Another way of seeing this, is that having Minkowski space in 
9-dimensions implies that the lifted solution is Minkowski space
in 10 dimensions (if the compactification radius is taken to be very large). 
If we then consider the setup of \cite{art:koedall}, namely $\lambda =i$, $m_{1}=m_{2}=0$
and $m_{3}=m\neq 0$, we can see that the lifted 10-dimensional killing spinor would have to
have the following $y$-dependence:
\begin{equation}
   \label{eq:KK_eis}
   \hat{\epsilon} (x,y) \;=\; e^{imy/4}\hat{\epsilon} (x) \; .
\end{equation} 
The killing spinors of Minkowski$_{10}$, in the Cartesian coordinates we use,
do not have any coordinate dependence whatsoever, so the only consistent
choice is for them to vanish. {\em I.e.} although Minkowski$_{10}$
gives rise to a solution of the gauged 9-dimensional supergravity, it
will not be supersymmetric.
\par
This point is comparable to the discussion in the literature of
why some supersymmetric solutions will give rise to non-supersymmetric
solutions after T-duality \cite{art:BKO,art:AAGB}. The most striking example
of this is Minkowski with an $\mathbb{R}^{2}$ written in polar coordinates, $r$ and $\varphi$.
Applying T-duality in the angular direction, it was found that the T-dual theory
was not supersymmetric, whereas the original obviously was maximally
supersymmetric. The reason is that in polar coordinates, IIB admits a Killing spinor  
\begin{equation}
 \label{eq:conic_susy}
 \hat{\epsilon} \;=\; \exp\left(  
                         -\textstyle{\frac{\varphi}{2}} \Gamma^{\underline{r}\underline{y}}
                    \right)\ \epsilon_{0}
\end{equation}
where $\epsilon_{0}$ is some arbitrary spacetime independent spinor
satisfying $\Gamma_{11}\epsilon_{0}= -\epsilon_{0}$, and $\underline{x}$
means the tangent-space direction associated to the coordinate $x$.
Now usual KK reduction prohibits any dependence on the compact coordinate,
so that the only consistent choice for $\epsilon_{0}$ is for it to vanish. 
Thus explaining how T-duality can connect a supersymmetric and
a non-supersymmetric solution.
\par
To conclude, we derived the supersymmetry transformation rules for the massive nine dimensional
supergravity. As this theory reduces in a certain limit to $N=2$ $d=9$ gauged supergravity, this permitted us
to show that the latter does not admit a supersymmetric Minkowski type of solution.
%The reason for this was that lifting the
%solution corresponds to a non-trivially $U(1)$ fibred Minkowski$_9$, which is not supersymmetric. 
%\par
%The above example shows the way, however. 
%We can diagonalize the $\Gamma$-matrices in Eq. (\ref{eq:conic_susy}), which
%causes the $\epsilon_{0}$ to be split into two parts $\epsilon^{+}_{0}$
%and $\epsilon_{0}^{-}$, defined by 
%$\Gamma^{\underline{r}\underline{y}}\epsilon^{\pm}_{0}\ =\ \pm i\epsilon^{\pm}_{0}$.
%Using this decomposition, the Killing spinor in Eq. (\ref{eq:conic_susy}) 
%can be written as
%\begin{equation}
% \label{eq:conic_susy2}
% \epsilon \;=\; \exp\left( -i\textstyle{\frac{Q}{2}}y\right)\epsilon_{0}^{+}
%                \,+\,
%                \exp\left( +i\textstyle{\frac{Q}{2}}y\right)\epsilon_{0}^{-}
%\end{equation}
%By choosing 
%Now, spinors on a circle are defined to satisfy $\Psi (x+2\beta )=\Psi (x)$,
%where $x$ is the coordinate on the circle and $\beta$ is its periodicity.
%Clearly, the above spinor satisfies this criterion when $y$ has period 
%$2\pi /Q$, which is the case when we have Minkowski$_{10}$ in Eq. (\ref{eq:conic}).
%%%%%%%%%%%%%%%%%%%%%%%%%%%%%%%%%%%%%%%%%%%%%%%%%%%%%%%%%%%%%%%%%%%%%%%%%%%%
%%%%%%%%%%%%%%%%%%%%%%%%%%%%%%%%%%%%%%%%%%%%%%%%%%%%%%%%%%%%%%%%%%%%%%%%%
%
%      M(assive)????
%
%%%%%%%%%%%%%%%%%%%%%%%%%%%%%%%%%%%%%%%%%%%%%%%%%%%%%%%%%%%%%%%%%%%%%%%%%
\section{Supersymmetry of Massive 11-dimensional Sugra}
\label{sec:M(assive)}
%%%%%%%%%%%%%%%%%%%%%%%%%%%%%%%%%%%%%%%%%%%%%%%%%%%%%%%%%%%%%%%%%%%%%%%%%
As was said in the introduction, massive 11-dimensional sugra was proposed
in order to have a way of obtaining massive type IIA sugra by means of
ordinary KK-reduction. The price to pay is the explicit occurrence
of killing vectors in the action, which is supported by the investigations
of the M-theory origin of massive D-branes \cite{art:yolanda,art:tomas}.
These vector fields do not change the field content of the `original' 11-dimenional sugra, as they are not
dynamical.
The occurrence of the killing vectors forces the introduction of mass
parameters, gathered into a symmetric mass-matrix $Q$, into the action. These masses 
are associated to the charges of
the KK9/M9-branes. The KK9-brane is the oxidation of the D8-brane, and 
a KK9-brane action was put forward in \cite{art:sato}.
\par
\cite{art:BvdS} discussed the solution corresponding
to the KK9-brane and put forward the relevant part of the supersymmetry
variation of the 11-dimensional gravitino, in order to discuss
the BPSness of the KK9-brane.
The proposed form however, cannot be straightforwardly generalized
to include more killing vectors. Therefore in the spirit of 
\cite{art:BLO,art:MO} we will propose a generalized variation of the 11-dimensional
gravitino and check that upon dimensional reduction it
matches with massive type IIA and the variations derived in 
Section \ref{sec:9d_susy}.
\par
%The bosonic part of the $D=11$ supergravity action is 
%given by\footnote{Here we scale $C_{(3)} \to (2\kappa )^{-1}C_{(3)}$ and 
%$\Psi \to \kappa^{-1}\Psi$.}
%\begin{equation}
%  \label{eq:M_action}
%  \mathcal{S}_{d=11}\;=\; \int d^{11}x\sqrt{g}\left[
%           R(g)\,-\,\textstyle{\frac{1}{2\cdot 4!}}G^{2}_{(4)}
%        \right]\,-\,\textstyle{\frac{1}{6}}\int_{11} dC_{(3)}\land dC_{(3)}\land C_{(3)} \; ,
%\end{equation}
%where $G_{(4)}\ =\ dC_{(3)}$. The supersymmetry variations are
%\begin{eqnarray}
%  \label{eq:M_SUSY_VAR}
%  \delta e_{\mu}{}^{a} &=& -i\bar{\epsilon}\Gamma^{a}\Psi_{\mu} \; , \nonumber \\
%  \delta \Psi_{\mu} &=& \nabla_{\mu}\epsilon \,+\, \textstyle{\frac{i}{288}}
%          \left[
%            \Gamma^{\alpha\beta\gamma\delta}{}_{\mu}\ -\
%            8 \Gamma^{\beta\gamma\delta}\delta_{\mu}{}^{\alpha}
%          \right]\epsilon F_{\alpha\beta\gamma\delta} \; , \nonumber \\
%  \delta C_{\alpha\beta\gamma} &=& 3\ \bar{\epsilon}\Gamma_{[\alpha\beta}\Psi_{\gamma ]} \; ,
%\end{eqnarray}
%Note that this means that we have rescaled the fields such that 
%the ever present $\kappa$ in the original papers vanishes.
%\par
As soon as we introduce more than one killing direction in our system,
the number of possible terms one can write down that are linear in 
the mass matrix and respect the tensor form of the massless variation,
is rather limited.
In fact the gravitino variation can be written as
\begin{eqnarray}
\delta\hat{\psi}_{\hat{\mu}} &=& 
   \nabla_{\hat{\mu}}\left(\check{\omega}\right)\hat{\epsilon}
\,+\,\textstyle{\frac{i}{288}}\left[
\hat{\Gamma}^{\hat{a}\hat{b}\hat{c}\hat{d}}{}_{\hat{\mu}}
\,-\, 8 \hat{\Gamma}^{\hat{b}\hat{c}\hat{d}}{\hat{e}_{\hat{\mu}}}^{\hat{a}}
\right] \hat{\epsilon} \hat{G}_{\hat{a}\hat{b}\hat{c}\hat{d}} \nonumber \\
& &-\,\textstyle{\frac{i}{12}}
     \hat{k}_{(n)\hat{\nu}}Q^{nm}\hat{k}_{(m)}^{\hat{\nu}}
        \hat{\Gamma}_{\hat{\mu}}\hat{\epsilon}
\,+\, \textstyle{\frac{i}{2}} \hat{k}_{(n)\hat{\mu}}Q^{nm}
       \hat{k}_{(m)\hat{\nu}}\hat{\Gamma}^{\hat{\nu}}\hat{\epsilon} \, ,
\label{eq:Msusy}
\end{eqnarray}
where $k^\mu_{(m)}$ are the $M$ killing vectors ($m=1,\dots,M$),  
\begin{equation}\label{fs}
\hat{G} \;=\; d\hat{C}\,-\,
\textstyle{\frac{1}{2}}i_{\hat{k}_{(n)}}\hat{C}
      Q^{nm}
       i_{\hat{k}_{(m)}}\hat{C} \; ,     
\label{eq:MG}
\end{equation}
and where the torsion, the contorsion and the connection are defined by
\begin{equation}
  \label{eq:Tor_def}
  T_{ab}{}^{c} \;=\; -\left[ i_{k_{(m)}}\hat{C}\right]_{ab} Q^{mn}k_{(n)}^{c}
  \hspace{.2cm},\hspace{.2cm}
  2 K_{abc} \;=\; -T_{abc}+T_{bca}-T_{cab} \; ,\;
  \check{\omega} \;=\; \omega + K \; .
 \end{equation}
Note that the definition of the field-strength (\ref{fs}) automatically
takes care of the fact that the field-strengths in lower dimensions
have the correct form \cite{art:BLO,art:MO}.
%%%%%%%%%%%%%%%%%%%%%%%%%%%%%%%%%%%%%%%%%%%%%%%%%%%%%%%%%%%%%%%%%%%%%%%%%%%%
\subsection{Towards Romans' theory}
%%%%%%%%%%%%%%%%%%%%%%%%%%%%%%%%%%%%%%%%%%%%%%%%%%%%%%%%%%%%%%%%%%%%%%%%%%%%
As was said before, the idea for massive 11-dimensional supergravity
was first introduced \cite{art:BLO}
in order to resolve the question of the 11-dimensional origin of 
Romans' theory. Therefore, the first thing to do would be to compare
the susy rules for Romans' theory \cite{art:romans} with the ones we
obtain from eq. (\ref{eq:Msusy}) by dimensional reduction over the killing direction,
which we will denote by $z$. As there is only one killing vector in this setting, we obviously have to take
$\hat{k}^{\hat{\mu}}\partial_{\hat{\mu}}\,=\, \partial_{z}$. 
In short, we decompose the bosonic fields as \cite{art:BLO}
\begin{equation}
\hat{C} \;=\; C\,+\, Bdz \hspace{.2cm},\hspace{.2cm}
\left(
   \hat{e}_{\hat{\mu}}{}^{\hat{a}}
\right)\;=\;
\left(
\begin{array}{ccc}
 e^{-\phi /3} e_{\mu}{}^{a} &\hspace{.2cm} & e^{2\phi /3}C^{(1)}_{\mu} \\
 & & \\
 0 & & e^{2\phi /3}
\end{array}
\right)\; .
\end{equation}
and the fermionic objects as
\begin{equation}
\begin{array}{lclclcl}
\hat{\psi}_{\underline{z}} &=& \textstyle{\frac{i}{3}}
                                  e^{\phi /6}\Gamma_{11}\lambda 
&\hspace{.5cm},\hspace{.5cm}&
\hat{\Gamma}^{a} &=& \Gamma^{a} \; , \\
& & & & & & \\
\hat{\psi}_{a} &=& e^{\phi /6}\left(
                         \psi_{a}\,-\,
                         \textstyle{\frac{1}{6}}\Gamma_{a}\lambda
                   \right)
&,& 
\hat{\Gamma^{10}} &=& i\hat{\Gamma}^{1\ldots 9}= -i\Gamma_{11} \; , \\
&& && && \\
\hat{\epsilon} &=& e^{-\phi /6}\epsilon \; , 
\end{array}
\end{equation}
where $a=0,\dots,9$, the tangent-space direction associated to $z$ is called $\underline{z}$
and $\epsilon$ is a 10-dimensional Majorana spinor.
\par
In accordance with our choice of $Q^{mn}$ in the next subsection, we take $Q=-m$ and can see that 
\begin{equation}
  \hat{k}^{a}\;=\; 0\; ,\;
  \hat{k}^{\underline{z}}\;=\; e^{2\phi /3} \hspace{.3cm}\to\hspace{.3cm}
  \hat{k}_{(n)\hat{\mu}}Q^{nm}\hat{k}_{(m)}^{\hat{\mu}}\;=\; m\ e^{4\phi /3} \; .
\end{equation}
Using the bosonic decomposition we can find
\begin{equation}
 \begin{array}{lclclcl}
  \hat{G}_{abcd} &=& e^{4\phi /3}G_{(4)abcd} &\hspace{.3cm},\hspace{.3cm}&
     G_{(4)} &=& dC_{(3)}-HC_{(1)}+\textstyle{\frac{m}{2}}B^{2} \; , \\
  \hat{G}_{abc\underline{z}} &=& e^{\phi /3}H_{abc} &,& H &=& dB \; ,\\
  \check{\omega}_{ab\underline{z}} &=& 
      \textstyle{\frac{1}{2}} e^{4\phi /3} G_{(2)ab} &,&
  G_{(2)} &=& dC_{(1)}+mB \; ,
 \end{array}
\end{equation}
and the rest of the objects as usual \cite{art:BLO}.
\par
Plugging everything into the supersymmetry variation (\ref{eq:Msusy}), we find
\begin{eqnarray}
  \delta\lambda &=& \slashed{\partial}\phi\epsilon 
        +\textstyle{\frac{1}{2\cdot 3!}}\Gamma_{11}\slashed{H}\epsilon
        +\textstyle{\frac{i}{4}}e^{\phi}
           \left\{
              5m 
             +\textstyle{\frac{3}{2}}\slashed{G}_{(2)} \left( -\Gamma_{11}\right)
             +\textstyle{\frac{1}{4!}}\slashed{G}_{(4)}
           \right\}\epsilon \; , \nonumber \\
 & & \nonumber \\
  \delta\Psi_{\mu} &=& \nabla_{\mu}\epsilon
          -\textstyle{\frac{1}{8}}\Gamma_{11} \slashed{H}_{\mu}\epsilon
          +\textstyle{\frac{i}{8}}e^{\phi}
             \left[
                m\Gamma_{\mu} \epsilon
               +\textstyle{\frac{1}{2}}\slashed{G}_{(2)}\Gamma_{\mu}
                      \left( -\Gamma_{11}\right)\epsilon
               +\textstyle{\frac{1}{4!}}\slashed{G}_{(4)\mu}\epsilon 
             \right] \; ,
\end{eqnarray}
which are, in our conventions, the string-frame supersymmetry variations of
massive type IIA (See {\em e.g.} \cite{art:JMO}). 
%%%%%%%%%%%%%%%%%%%%%%%%%%%%%%%%%%%%%%%%%%%%%%%%%%%%%%%%%%%%%%%%%%%%%%%%%%%%
\subsection{Comparison with the results in Sec \ref{sec:9d_susy}}
%%%%%%%%%%%%%%%%%%%%%%%%%%%%%%%%%%%%%%%%%%%%%%%%%%%%%%%%%%%%%%%%%%%%%%%%%%%%
{}We now reduce (\ref{eq:Msusy}) to nine dimensions on a two-torus (using regular Kaluza-Klein truncation) and check if we find the tranformation rules for the
massive nine dimensional supergravity. Following the conventions of \cite{art:MO} our Ansatz for the metric is
\begin{equation}
  \label{eq:M2d9}
  \hat{e}_{\hat{\mu}}{}^{\hat{a}}\,=\,
    \left(
      \begin{array}{cc}
        K^{-1/7}\ e_{\mu}{}^{a} & K^{1/2}\ v_{m}{}^{i} A^{(m)}_{\mu} \\
        0 & K^{1/2}\ v_{m}^{i}
      \end{array}
    \right)\; ,\;
  \hat{e}_{\hat{a}}{}^{\hat{\mu}}\,=\,
    \left(
      \begin{array}{cc}
        K^{1/7}\ e_{a}{}^{\mu} & - K^{1/7}\  A^{(m)}_{a} \\
        0 & K^{-1/2}\ v_{i}^{m}
      \end{array}
    \right)\; 
\end{equation}
where $i,j,m,n=1,2$, $K= \left( \det \hat{g}_{mn}\right)^{1/2}$ and $v_{m}{}^{i}v_{n}{}^{j}\delta_{ij}=\mathcal{M}_{mn}$.
Or, put differently, $\mathcal{M}=VV^{T}$ with $V= v_{m}{}^{i}$. 
Since $V$ is an element of $Sl(2,\mathbb{R})$, we also have
$V^{-1}=\varepsilon V^{T}\varepsilon^{T}=v_{i}{}^{m}$.
Looking at (\cite{art:MO}:2.15) we see that
\begin{equation}
  v_{m}{}^{i}\;=\;\left(
      \begin{array}{cc}
        e^{-\phi /2} & e^{\phi /2}C_{(0)}\\
        0            & e^{\phi /2}
      \end{array}
   \right) \hspace{.3cm},\hspace{.3cm}
  v_{i}{}^{m}\;=\;
   \left(
     \begin{array}{cc}
         e^{\phi /2} & - e^{\phi /2}C_{(0)}\\
         0           & e^{-\phi /2}
     \end{array}
   \right) \; .
\end{equation}
And as before we have $A_{\mu}^{(m)} =-\varepsilon^{mn}A_{\mu (n)}$. Doublets of fields like $A_\mu^{(m)}$
will be written as vectors (like $\vec{A}_\mu$).
The 3-form will be split as
\begin{equation}
  \label{eq:M-9_3form}
  \hat{C}\;=\; C_{(3)}
    -\textstyle{\frac{1}{2}} \vec{A}^{T}_{(1)}\varepsilon\vec{A}_{(2)}
    +\vec{A}_{(2)}^{T}\vec{\theta}
    +\textstyle{\frac{1}{2}}A_{(1)}\vec{\theta}^{T}\varepsilon\vec{\theta} \; ,
\end{equation}
where we defined $\vec{\theta}=d\vec{y}-\varepsilon\vec{A}_{(1)}$.
A small calculation then immediately shows that
\begin{equation}
  \begin{array}{{lclclcl}}
    \hat{G}_{abij} &=& K^{5/7}F_{(2)ab}\varepsilon_{ij} &\; ,\;&
        F_{(2)} &=& dA_{(1)} \; , \\
    \hat{G}_{abci} &=& K^{-1/14}v_{i}{}^{m} F_{(3)abcm} &,&
      \vec{F}_{(3)} &=& d\vec{A}_{(2)}\,+\, A_{(1)}\vec{F}_{(2)} \; ,\\
    \hat{G}_{abcd} &=& K^{4/7}G_{(4)abcd} &,& 
    G_{(4)} &=& dC_{(3)}
            +\textstyle{\frac{1}{2}}\vec{A}_{(1)}^{T}\varepsilon d\vec{A}_{(2)}
%            +\textstyle{\frac{1}{2}}\vec{A}_{(1)}^{T}\varepsilon\vec{F}_{(3)}
            -\textstyle{\frac{1}{2}}\vec{A}_{(2)}^{T}\varepsilon\vec{F}_{(2)} \; ,\\
%            +\textstyle{\frac{1}{2}}A_{(1)}\vec{A}_{(1)}^{T}\varepsilon\vec{F}_{(2)} \; ,\\
%
    \check{\omega}_{abi} &=& \textstyle{\frac{1}{2}}K^{11/14}v_{im}\varepsilon^{mn} F_{(2)abn} &,&
       \vec{F}_{(2)} &=& d\vec{A}_{(1)}-M\vec{A}_{(2)} \; , \\
    \check{\omega}_{aij} &=& -K^{1/7}v_{[i|}{}^{m}\mathcal{D}_{a}v_{m|j]} &,&
       \mathcal{D}v_{mj} &=& dv_{mj}\,-\, A_{(1)} M_m{}^{n}v_{nj} \; , \\
    &&&&\check{\omega}_{iaj} &=& K^{1/7}\left[
                           \textstyle{\frac{1}{2}}\eta_{ij}\partial_{a}\log (K)
                          +v_{(i|}{}^{m}\mathcal{D}_{a}v_{m|j)}
                        \right] \; .   
  \end{array}
\end{equation}
where $\eta_{ij}=-\delta_{ij}$.
\par
The Clifford algebra will be split as
\begin{equation}
  \Gamma^{a} \;=\; \sigma^{2}\otimes\gamma^{a}\; ,\;
  \Gamma^{9} \;=\; i\sigma^{1}\otimes \mathbb{I}\; ,\;
  \Gamma^{10} \;=\; -i\sigma^{3}\otimes\mathbb{I} \; ,
\end{equation}
%With this we can see that $\Gamma^{ij} = -i\bar{\varepsilon}^{ij} \sigma^{2}$.
and the Ansatz for the 11-dimensional gravitino is \cite{art:KenK}
\begin{equation}
  \hat{\Psi}_{i}\;=\; K^{1/14} \chi_{i}\; ,\,
  \hat{\Psi}_{a}\;=\; \ K^{1/14}\left\{
                        \Psi_{a}\,-\,
                        \textstyle{\frac{1}{7}}\Gamma_{a}\Gamma^{i}\chi_{i}
                      \right\} \;,\;
  \hat{\epsilon}\;=\; K^{-1/14} \epsilon \; ,
\end{equation}
which means that we combine the 9-dimensional spinors into a vector. 
%\par
%The needed Lorentz transformations are
%\begin{equation}
%  \omega^{a}{}_{i}\;=\; -i \bar{\hat{\epsilon}}\Gamma^{a}\hat{\Psi}_{i}\;,\;
%  \omega^{ab}\;=\; \textstyle{\frac{i}{??}}\bar{\hat{\epsilon}}
%                         \Gamma^{ab}\Gamma^{i}\hat{\Psi}_{i} \; .
%\end{equation}
After a small calculation we then end up with
\begin{eqnarray}
%  \delta\ e_{\mu}{}^{a} &=& \bar{\epsilon}\gamma^{a}\Psi_{\mu} \\
%%
%  \delta\ \log\left( K\right) &=& 
%       \bar{\epsilon}\sigma^{2}\Gamma^{i}\chi_{i}  \\
%%
  \delta\ \chi_{i} &=& 
       -\textstyle{\frac{1}{4}}\slashed{\partial}\log\left( K\right)
           \sigma^{2}\Gamma_{i}\epsilon
       -\textstyle{\frac{1}{2}} v_{(i|}{}^{m}\slashed{\mathcal{D}}v_{m|j)} \sigma^{2}\Gamma^{j}\epsilon
       +\textstyle{\frac{i}{12\cdot 4!}}K^{3/7}
           \slashed{F}_{(4)}\Gamma_{i}\epsilon \nonumber \\
 & & \nonumber \\
  & &
       -\textstyle{\frac{1}{8}}K^{9/14}v_{mi}\slashed{F}_{(2)}^{m}\epsilon
       +\textstyle{\frac{i}{12\cdot 3!}}K^{-3/14}
            v_{j}{}^{m}\slashed{F}_{(3)m}\sigma^{2}\Gamma^{j}{}_{i}\epsilon
       +\textstyle{\frac{i}{6\cdot 3!}}K^{-3/14}
            v_{i}{}^{m}\slashed{F}_{(3)m} \sigma^{2}\epsilon
       \nonumber \\
  & & \nonumber \\
  & &
       -\textstyle{\frac{i}{2\cdot 3!}}K^{-6/7}\slashed{F}_{(2)}
            \varepsilon_{ij}\Gamma^{j}\epsilon 
       +\textstyle{\frac{i}{12}}K^{6/7}Tr\left(Q\mathcal{M}\right)
            \hat{\Gamma}_{i}\epsilon
       +\textstyle{\frac{i}{2}}K^{6/7}v_{ni}Q^{nm}v_{mj}\hat{\Gamma}^{j}\epsilon \; , \label{eq:M_dilatino}\\
 & & \nonumber \\
  \delta\Psi_{\mu} &=&
       \mathfrak
{D}_{\mu}\epsilon
      +\textstyle{\frac{i}{4\cdot 7}}K^{6/7} Tr\left( Q\mathcal{M}\right)
           \gamma_{\mu}\epsilon
      -\textstyle{\frac{K^{9/14}}{7\cdot 8}}v_{mi} F_{(2)ab}^{m}
           \left[
              \gamma_{\mu}{}^{ab}-12 e_{\mu}{}^{a}\gamma^{b}
           \right] \sigma^{2}\hat{\Gamma}^{i}\epsilon  % \nonumber \\
   \nonumber \\
 & & \nonumber \\
  & & 
      +\textstyle{\frac{iK^{-6/7}}{7\cdot 16}} F_{(2)ab}
           \left[
              \gamma_{\mu}{}^{ab}-12 e_{\mu}{}^{a}\gamma^{b}
           \right] \sigma^{2}\varepsilon_{ij}\hat{\Gamma}^{ij}\epsilon
           +\textstyle{\frac{iK^{3/7}}{4\cdot 7\cdot 4!}}F_{abcd}
           \left[
              3\gamma_{\mu}{}^{abcd}-16 e_{\mu}{}^{a}\gamma^{bcd}
           \right]\sigma^{2}\epsilon
        % \nonumber \\
% 
 \nonumber \\
 & & \nonumber \\
  & & +\textstyle{\frac{iK^{-3/14}}{4\cdot 7\cdot 3!}} v_{i}{}^{m} F_{(3)abcm}
           \left[
              2\gamma_{\mu}{}^{abc} -15 e_{\mu}{}^{a}\gamma^{bc}
           \right] \hat{\Gamma}^{i}\epsilon
    \; , \label{eq:M_gravitino} 
\end{eqnarray}
where we have defined
\begin{equation}
  \mathfrak{D}\epsilon \,=\, \nabla\epsilon\,+\,
       \textstyle{\frac{1}{4}} v_{i}{}^{m}\mathcal{D}v_{mj}
            \Gamma^{ij}\epsilon \; =\;
  \nabla\epsilon\,-\,\textstyle{\frac{i}{2}}\mathcal{Q}_{(1)} \sigma^{2} \epsilon \; ,
\end{equation}
{}For the last step in the above identification, we made use of the fact that 
\begin{equation}
  Tr\left( Q\mathcal{M}\right) \;=\; 
       2i\vec{V}_{-}^{T}\varepsilon M\vec{V}_{+} \,=\,
       \frac{e^{\phi}}{2}\left(
         m_{3}+m_{2}
         +2m_{1}C_{(0)}
         +(m_{3}-m_{2})\bar{\lambda}\lambda
       \right) \; . 
\end{equation}
Note that when we consider the case with $Q=0$, 
Eqs. (\ref{eq:M_dilatino},\ref{eq:M_gravitino}) reduce to the expressions
given in \cite{art:KenK} for $N=2$, $d=9$ supergravity.
\par
By identifying 
\begin{equation}
\epsilon \;=\; \left(\begin{array}{c} 
                     \epsilon_{1} \\ -\epsilon_{2}
               \end{array}\right) \;,\;
\Psi_{\mu} \;=\; \left(\begin{array}{c} 
                     \Psi_{1\mu} \\ -\Psi_{2\mu}
               \end{array}\right) \;,\;
\tilde{\chi} \;=\; \left(\begin{array}{c} 
                     \tilde{\chi}_{1} \\ -\tilde{\chi}_{2}
               \end{array}\right) \;,\;
\chi \;=\; \left(\begin{array}{c} 
                     \chi_{1} \\ \chi_{2}
               \end{array}\right) \;,
\end{equation}
introducing the combination 
$\chi_{ij} \;=\; \sigma^{2} \Gamma_{(i}\chi_{j)}-\textstyle{\frac{1}{2}}\sigma^{2} \eta_{ij} \Gamma^{l} \chi_{l}$
and introducing the complex combinations $\xi = \xi_{1}+i\xi_{2}$, {\em e.g.} 
$\epsilon =\epsilon_{1}+i\epsilon_{2}$, we can achieve a complete correspondence if we take
\begin{equation}
 \label{eq:correspondence}
 \tilde{\chi}\;=\; -\textstyle{\frac{3}{4}}\sigma^{2} \Gamma^{i}\chi_{i} \hspace{.3cm},\hspace{.3cm}
 \chi \;=\; \chi_{11} \,+\, i\sigma^{2}\chi_{12} \; .
\end{equation}
Thus, regular Kaluza-Klein reduction on a two-torus of massive 11-dimensional supergravity gives rise to the massive 
$d=9$ supergravity.
%%%%%%%%%%%%%%%%%%%%%%%%%%%%%%%%%%%%%%%%%%%%%%%%%%%%%%%%%%%%%%%%%%%%%%%%%%%%
%
%   Conclusies... Indien...
%
%%%%%%%%%%%%%%%%%%%%%%%%%%%%%%%%%%%%%%%%%%%%%%%%%%%%%%%%%%%%%%%%%%%%%%%%%%%%
\section{Conclusions}
In this letter we have reduced the IIB supergravity in a
generalized way as to find the supersymmetry variations
of the $Sl(2,\mathbb{R})$ family of 9-dimensional supergravity
theories presented in \cite{art:LLP,art:MO}.
This enabled us to find the supersymmetry variations
of the 9-dimensional gauged supergravity, proposed recently
in \cite{art:koedall}.
Contrary to the claims made in \cite{art:koedall} 9-dimensional
Minkowski space is not a supersymmetric solution of the 
9-dimensional $SO(2)$ gauged supergravity, mainly because the 
consistency of the KK-Ansatz on the killing spinors requires them
to vanish. 
\par
We gave the possible form of the gravitino variations of
what is called massive 11-dimensional supergravity and have
shown that it gives rise to the correct result when compared to 
Romans' theory and the supersymmetry variations we obtained
by generalized dimensional reduction from type IIB.
\par
Needless to say, there still is a long way to go. It would be nice
to understand the underlying structure of the 11-dimensional massive
supergravity, and use this structure to derive the complete 11-dimensional
supergravity. It would also be nice to see if the 11-dimensional supergravity
when compactified to lower dimensions will give
rise to known theories or new theories which are related to the known
ones by some S-duality relation as happens in \cite{art:AMO}.
Surely, the question whether 11-dimensional massive supergravity is a 
consistent construct, which might be called M(assive) theory, 
deserves further investigation.
%%%%%%%%%%%%%%%%%%%%%%%%%%%%%%%%%%%%%%%%%%%%%%%%%%%%%%%%%%%%%%%%%%%%%%%%%%%%
%
%   de bedankjes...
%
%%%%%%%%%%%%%%%%%%%%%%%%%%%%%%%%%%%%%%%%%%%%%%%%%%%%%%%%%%%%%%%%%%%%%%%%%%%%
\section*{Acknowledgments}
The authors would like to thank T. Ort\'{\i}n and A. van Proeyen for fruitful discussions
and comments.
This work was supported in part by the F.W.O.-Vlaanderen, and 
the E.U. RTN programmes HPRN-CT-2000-00131 and HPRN-CT-2000-00148.
%%%%%%%%%%%%%%%%%%%%%%%%%%%%%%%%%%%%%%%%%%%%%%%%%%%%%%%%%%%%%%%%%%%%%%%%%%%%
%
%     Appendices.... Indien nodig...
%
%%%%%%%%%%%%%%%%%%%%%%%%%%%%%%%%%%%%%%%%%%%%%%%%%%%%%%%%%%%%%%%%%%%%%%%%%%%%
\appendix{
%%%%%%%%%%%%%%%%%%%%%%%
%%%%%%%%%%%%%%%%%%
\section{From $Sl(2,\mathbb{R})$ to $SU(1,1)$ and back}
\label{app:groups}
%%%%%%%%%%%%%%%%%%
A defining matrix in $Sl(2,\mathbb{R})$, $V$, and $SU(1,1)$, $\mathcal{V}$, are parametrized by
\begin{equation}
  \label{eq:Element_Def}
\left.
\begin{array}{c}
  V \,=\, \left(
    \begin{array}{cc}
       a & b \\
       c & d
    \end{array}
  \right) \\[.4cm]
 ad-bc\,=\, 1
\end{array}
\right\}\,\in\, Sl(2,\mathbb{R}) \hspace{.5cm},\hspace{.5cm}
\left.
\begin{array}{c}
  \mathcal{V} \,=\, \left(
    \begin{array}{cc}
       u & v \\
       \bar{v} & \bar{u}
    \end{array}
  \right)\\[.4cm]
  u\bar{u}-v\bar{v}\,=\, 1
\end{array}
\right\} \,\in\, SU(1,1) \; .
\end{equation}
The explicit mapping from $Sl(2,\mathbb{R})$ to $SU(1,1)$ can be found to be 
\begin{equation}
u \,=\, \frac{d+a}{2}\;+\; i\frac{b-c}{2} \hspace{.2cm},\hspace{.2cm}
v \,=\, \frac{d-a}{2}\;+\; i\frac{b+c}{2} \; ,
\label{eq:Group_Map}
\end{equation}
which obviously can be inverted.
Splitting the $SU(1,1)$ matrix into two vectors as $\mathcal{V}= (\vec{\mathcal{V}}_{-}\vec{\mathcal{V}}_{+})$,
we parametrize the coset $SU(1,1)/U(1)$ as
\begin{equation}
  \label{eq:SU11_coset_param}
  \vec{\mathcal{V}}_{+} \;=\; \frac{1}{2Im(\lambda )^{1/2}}\left(
            \begin{array}{c}
              1\,+\, i\lambda \\
              1\,-\, i\lambda
            \end{array}
          \right) \hspace{.4cm},\hspace{.4cm} 
  \vec{\mathcal{V}}_{-}\,=\, \sigma^{1}\vec{\mathcal{V}}_{+}^{*}
 \; ,
\end{equation}
with $\lambda = C_{(0)}+ie^{-\phi}$ the so-called axidilaton.
In order to stay in our choice of parameterization,
an $SU(1,1)$-transformation has to be accompanied
by a local $U(1)$-transformation, {\em i.e.} the $SU(1,1)$-transformation is
$\vec{\mathcal{V}}_{\pm}=e^{\pm i\Sigma}\Lambda\ \vec{\mathcal{V}}_{\pm}^{\prime}$.
The parameter $\Sigma (x)$ is readily found to be defined by
\begin{equation}
  \label{eq:Comp_Trans}
  e^{2i\Sigma} \;=\; \frac{c\bar{\lambda}+d}{c\lambda +d} \; ,
\end{equation}
The other $SU(1,1)$ fields are given in terms of their $Sl(2,\mathbb{R})$
counterparts by
\begin{equation}
  \label{eq:2-forms}
  D_{(4)} \,=\, C_{(4)}\ -\ \textstyle{\frac{1}{2}}B\land C_{(2)} \; ,\;
  \mathcal{A}_{(2)}^{1} \;=\; B\,+\,i\ C_{(2)}\hspace{.5cm},\hspace{.5cm}
  \mathcal{A}_{(2)}^{2} \;=\; \left( \mathcal{A}_{(2)}^{1} \right)^{*} \; .
\end{equation}
where $C_{(4)}$ is the usual string field. 
With the above identifications one finds
\begin{eqnarray}
  \mathcal{P}_{(1)} &=& \textstyle{\frac{1}{2}}d\phi\,+\,
                  \textstyle{\frac{i}{2}}e^{\phi}dC_{(0)} \; \\
  \mathcal{Q}_{(1)} &=& -\textstyle{\frac{1}{2}}e^{\phi}dC_{(0)} \; ,\\
  \mathcal{G}_{(3)} &=& e^{-\phi /2}H\,+\, i\ e^{\phi /2}G_{(3)} \; , 
\end{eqnarray}
where $G_{(3)}=dC_{(2)}-H\ C_{(0)}$ is the standard field-strength
for the RR 2-form.
}
%%%%%%%%%%%%%%%%%%%%%%%%%%%%%%%%%%%%%%%%%%%%%%%%%%%%%%%%%%%%%%%%%%%%%%%%%%%
%
%    Referenties...
%
%%%%%%%%%%%%%%%%%%%%%%%%%%%%%%%%%%%%%%%%%%%%%%%%%%%%%%%%%%%%%%%%%%%%%%%%%%%

%%%%%%%%%%%%%%%%%%%%%%%%%%%%%%%%%%%%%%%%%%%%%%%%%%%%%%%%%%%%%%%%%%%%%%%%%%
%
%   That's all folks!!!!!!!!!!
%
%%%%%%%%%%%%%%%%%%%%%%%%%%%%%%%%%%%%%%%%%%%%%%%%%%%%%%%%%%%%%%%%%%%%%%%%%%
\end{document}